\newcommand{\std}[1]{
  \usepackage{chicago}
  \usepackage{amssymb}
  \usepackage{amsmath}
  \usepackage[final]{graphicx}
  \usepackage{a4,a4wide}
  \renewcommand{\baselinestretch}{#1}
  \renewcommand{\arraystretch}{1.2}
  \parindent 3ex
  \parskip 1ex
  \newcommand{\blockindent}{3ex}
  \pagestyle{headings}
  \macros
  \begin{document}
}
\newcommand{\article}[2]{
  \documentclass[#1pt,fleqn]{article}\std{#2}
}
\newcommand{\docu}[1]{
  \documentclass[12pt,fleqn]{article}\std{#1}
  \usepackage{Lweb}
}
\newcommand{\artikel}[1]{
  \documentclass[12pt,twoside,fleqn]{article}\usepackage{german}\std{#1}
}
\newcommand{\revtex}{
  \documentstyle[preprint,aps,eqsecnum,amssymb,fleqn]{revtex}
  \begin{document}
}
\newcommand{\book}[1]{
  \documentclass[12pt,twoside,fleqn]{book}
  \addtolength{\oddsidemargin}{2mm}
  \addtolength{\evensidemargin}{-4mm}
  \std{#1}
}
\newcommand{\buch}[1]{
  \documentclass[10pt,twoside,fleqn]{book}
  \usepackage{german}
  \usepackage{makeidx}
  \makeindex
  \std{#1}
}
\newcommand{\foils}[2]{
  \documentclass[fleqn]{article}
  \usepackage{amssymb}
  \usepackage{amsmath}
  \usepackage[final]{graphicx}
  \renewcommand{\baselinestretch}{#1}
  \renewcommand{\arraystretch}{1.5}
  \setlength{\voffset}{-3cm}
  \setlength{\hoffset}{-4cm}
  \setlength{\textheight}{27cm}
  \setlength{\textwidth}{19cm}
  \parindent 0ex
  \parskip #2ex
  \pagestyle{plain}
  \begin{document}
  \huge
}
\newcommand{\landfoils}[1]{
  \documentclass[fleqn]{article}
  \usepackage{amssymb}
  \usepackage{amsmath}
  \usepackage[final]{graphicx}
  \renewcommand{\baselinestretch}{#1}
  \renewcommand{\arraystretch}{1.5}
  \setlength{\hoffset}{-5cm}
  \setlength{\voffset}{-1.5cm}
  \setlength{\textwidth}{27cm}
  \setlength{\textheight}{19cm}
  \parindent 0ex
  \parskip 0ex 
  \pagestyle{plain}
  \begin{document}
  \huge
}
\newcommand{\landfolien}[1]{
  \documentclass[fleqn]{article}
  \usepackage{german}
  \usepackage{amssymb}
  \usepackage{amsmath}
  \usepackage[final]{graphicx}
  \renewcommand{\baselinestretch}{#1}
  \renewcommand{\arraystretch}{1.5}
  \setlength{\hoffset}{-5cm}
  \setlength{\voffset}{-1.5cm}
  \setlength{\textwidth}{27cm}
  \setlength{\textheight}{19cm}
  \parindent 0ex
  \parskip 0ex 
  \pagestyle{plain}
  \begin{document}
  \huge
}


\newcommand{\horline}{
\vspace{-1ex}
\begin{list}{}{\leftmargin0ex \topsep0ex}\item[]\hrulefill\end{list}
\vspace{1ex}}

\renewcommand{\title}[1]{
\horline
\begin{list}{}{\leftmargin2ex \rightmargin5ex \topsep0ex }\item[]
\vspace{6ex}
{\huge\bf #1}
\vspace{4ex}
\begin{list}{}{\leftmargin7ex}\item[]
{\bf Marc Toussaint}\\
Institut f\"ur Neuroinformatik\\
Ruhr-Universit\"at Bochum, ND 04\\
44780 Bochum - Germany\\
{\tt www.neuroinformatik.ruhr-uni-bochum.de/PEOPLE/mt/}
\end{list}
\end{list}
\horline
}

\newcounter{parac}
\newcommand{\para}{\refstepcounter{parac}{\bf [{\roman{parac}}]}~~}
\newcommand{\Pref}[1]{[\emph{\ref{#1}}\,]}

\newcommand{\headline}[1]{\footnotetext{\sc Marc Toussaint, \today
    \hspace{\fill} file: #1}}

\newcommand{\sepline}{
\begin{center} \begin{picture}(200,0)
  \line(1,0){200}
\end{picture}\end{center}
}

\newcommand{\intro}[1]{\textbf{#1}\index{#1}}

\newtheorem{definition}{Definition}

\newenvironment{block}{
\begin{list}{}{\leftmargin\blockindent \topsep-\parskip}
\item[]
}{
\end{list}
}

\newenvironment{rblock}{
\begin{list}{}{\leftmargin\blockindent \rightmargin\blockindent \topsep-\parskip}
\item[]
}{
\end{list}
}

\newenvironment{abstrac}
{\paragraph{Abstract}\begin{rblock}\small}
{\end{rblock}}
\newenvironment{keywords}
{\paragraph{Keywords}\begin{rblock}\small}
{\end{rblock}}

\newenvironment{stat}[1]{
\begin{block}$\bullet\!\!\!$ 
}{
\end{block}
}

\newenvironment{enum}{
\begin{list}{}{\leftmargin3ex \topsep0ex \itemsep0ex}
\item[\labelenumi]
}{
\end{list}
}

\newenvironment{cramp}{
\begin{quote} \begin{picture}(0,0)
        \put(-5,0){\line(1,0){20}}
        \put(-5,0){\line(0,-1){20}}
\end{picture}

}{

\begin{picture}(0,0)
        \put(-5,5){\line(1,0){20}}
        \put(-5,5){\line(0,1){20}}
\end{picture} \end{quote}
}

\newenvironment{summary}{
\begin{center}\begin{tabular}{|l|}
\hline
}{\\
\hline
\end{tabular}\end{center}
}

\newcommand{\inputReduce}[1]{
  
  {\sc\hspace{\fill} REDUCE file: #1}
}
\newcommand{\inputReduceInput}[1]{
  
  {\sc\hspace{\fill} REDUCE input - file: #1}
  \input{#1.tex}
}
\newcommand{\inputReduceOutput}[1]{
  
  {\sc\hspace{\fill} REDUCE output - file: #1}
}

\newcommand{\macros}{
  \newcommand{\0}{{\hat 0}}
  \newcommand{\1}{{\hat 1}}
  \newcommand{\2}{{\hat 2}}
  \newcommand{\3}{{\hat 3}}
  \newcommand{\5}{{\hat 5}}
  \newcommand{\QQ}{{\cal Q}}

  \renewcommand{\a}{\alpha}
  \renewcommand{\b}{\beta}
  \renewcommand{\c}{\gamma}
  \renewcommand{\d}{\delta}
    \newcommand{\D}{\Delta}
    \newcommand{\e}{\epsilon}
    \newcommand{\g}{\gamma}
    \newcommand{\G}{\Gamma}
  \renewcommand{\l}{\lambda}
  \renewcommand{\L}{\Lambda}
    \newcommand{\m}{\mu}
    \newcommand{\n}{\nu}
    \newcommand{\N}{\nabla}
  \renewcommand{\k}{\kappa}
  \renewcommand{\o}{\omega}
  \renewcommand{\O}{\Omega}
    \newcommand{\p}{\varphi}
  \renewcommand{\P}{\Phi}
  \renewcommand{\r}{\varrho}
    \newcommand{\s}{\sigma}
    \newcommand{\Si}{\Sigma}
  \renewcommand{\t}{\theta}
    \newcommand{\T}{\Theta}
  \renewcommand{\v}{\vartheta}
    \newcommand{\Y}{\Upsilon}

  \newcommand{\C}{{\bf C}}
  \newcommand{\R}{{\bf R}}
  \newcommand{\Z}{{\bf Z}}

  \renewcommand{\AA}{{\cal A}}
  \newcommand{\GG}{{\cal G}}
  \renewcommand{\SS}{{\cal S}}
  \newcommand{\TT}{{\cal T}}
  \newcommand{\EE}{{\cal E}}
  \newcommand{\HH}{{\cal H}}
  \newcommand{\II}{{\cal I}}
  \newcommand{\KK}{{\cal K}}
  \newcommand{\MM}{{\cal M}}
  \newcommand{\NN}{{\cal N}}
  \newcommand{\CC}{{\cal C}}
  \newcommand{\PP}{{\cal P}}
  \newcommand{\RR}{{\cal R}}
  \newcommand{\YY}{{\cal Y}}
  \newcommand{\SOSO}{{\cal SO}}
  \newcommand{\GLGL}{{\cal GL}}

  \newcommand{\NNN}{\mathbb{N}}
  \newcommand{\ZZZ}{\mathbb{Z}}
  \newcommand{\RRR}{\mathbb{R}}
  \newcommand{\CCC}{\mathbb{C}}
  \newcommand{\one}{{\bf 1}}

  \newcommand{\<}{\langle}
  \renewcommand{\>}{\rangle}
  \newcommand{\cor}{{\rm cor}}
  \newcommand{\cov}{{\rm cov}}
  \newcommand{\sd}{{\rm sd}}
  \newcommand{\tr}{{\rm tr}}
  \newcommand{\lag}{\mathcal{L}}
  \newcommand{\inn}{\rfloor}
  \newcommand{\lie}{\pounds}
  \newcommand{\speer}{\parbox{0.4ex}{\raisebox{0.8ex}{$\nearrow$}}}
  \renewcommand{\dag}{ {}^\dagger }
  \newcommand{\h}{{}^\star}
  \newcommand{\w}{\wedge}
  \newcommand{\ow}{\stackrel{\circ}\wedge}
  \newcommand{\feed}{\nonumber \\}
  \newcommand{\comma}{\; , \quad}
  \newcommand{\period}{\; . \quad}
  \newcommand{\del}{\partial}
  \newcommand{\point}{$\bullet~~$}
  \newcommand{\doubletilde}{
  ~ \raisebox{0.3ex}{$\widetilde {}$} \raisebox{0.6ex}{$\widetilde {}$} \!\!
  }
  \newcommand{\topcirc}{\parbox{0ex}{~\raisebox{2.5ex}{${}^\circ$}}}
  \newcommand{\sym}{\topcirc}

  \newcommand{\half}{\frac{1}{2}}
  \newcommand{\third}{\frac{1}{3}}
  \newcommand{\fourth}{\frac{1}{4}}

  \renewcommand{\small}{\footnotesize}
}

\newcommand{\tmp}{\fbox{?}}
\newcommand{\Label}[1]{\label{#1}\fbox{\tiny #1}}

\newcommand{\path}{./}
\newcommand{\basepath}{./}
\newcommand{\setpath}[1]{
  \renewcommand{\path}{#1}
  \renewcommand{\basepath}{#1}}
\newcommand{\pathinput}[2]{
  \renewcommand{\path}{\basepath #1}
  \input{\path #2} \renewcommand{\path}{\basepath}}

\newcommand{\hide}[1]{[\small #1 \normalsize]}

\article{11}{1.1}

\renewcommand{\labelenumi}{\textbf{(\roman{enumi})}}

\renewcommand{\intro}[1]{\textbf{#1}}
\newcommand{\refer}[1]{[\emph{#1}]}

\thispagestyle{empty}

\title{Self-adaptive exploration in evolutionary search}

\begin{abstrac}
  We address a primary question of computational as well as biological
  research on evolution: \emph{How can an exploration strategy adapt
    in such a way as to exploit the information gained about the
    problem at hand?} We first introduce an integrated formalism of
  evolutionary search which provides a unified view on different
  specific approaches. On this basis we discuss the implications of
  indirect modeling (via a ``genotype-phenotype mapping'') on the
  exploration strategy. Notions such as modularity, pleiotropy and
  functional phenotypic complex are discussed as implications.

  Then, rigorously reflecting the notion of self-adaptability, we
  introduce a new definition that captures self-adaptability of
  exploration: different genotypes that map to the same phenotype may
  represent (also topologically) different exploration strategies;
  self-adaptability requires a variation of exploration strategies
  along such a ``neutral space''. By this definition, the concept of
  neutrality becomes a central concern of this paper.

  Finally, we present examples of these concepts: For a specific
  grammar-type encoding, we observe a large variability of exploration
  strategies for a fixed phenotype, and a self-adaptive drift towards
  short representations with highly structured exploration strategy
  that matches the ``problem's structure''.
\end{abstrac}



\section{Introduction}

Typically, when a problem is given, the space of all potential
solutions is too large to try all of them in reasonable time. If not
making \emph{any} further assumptions on the problem, there neither
exists a preferable strategy to search for solutions. Usually though,
one assumes that the problem is not notoriously arbitrary, that it has
some ``structure'' and that there might exist some smart strategies to
explore the space. More specifically, one hopes that one can draw
information from the quality of previously explored solutions on how
to choose new explorations. For example, when assuming some
``continuity''\footnote{which requires to identify a topology on the
  search space} of the problem, one may search further in regions of
previously explored good solutions.

A more elaborated strategy is the following: Analyze the statistics of
previously found solutions, find correlations between certain
characters (parameters) of the solution and the solution's quality,
find mutual information between the characters of good solutions,
etc., and exploit all this information to choose further explorations
--- in the hope that these findings really characterize the problem,
that the problem is characterizable by such information. In essence,
the latter approach will explore only a tiny part of $P$, strongly
dependent on early explorations that have been successful. Found
solutions may lay no claim to be globally optimal; they are a further
development of early successful concepts.

The central questions become: \emph{How can we analyze the statistics
  of previously explored and evaluated solutions? How can we represent
  this gained information? How can we model an exploration strategy
  depending on this information?}

One direct approach to these questions leads to statistical models of
exploration. For example, a Bayesian network can encode the
probability of future explorations (the \emph{exploration density})
and is trained with previously successful solution parameters (as done
by \citeN{pelikan:00}, see appendix \ref{ExpMod}). In contrast, we
will argue that the exploration strategy can be modeled by a mapping
onto the solution space, a genotype-phenotype mapping. This means that
a (simple) density on a base space (genotype) is lifted to the
exploration density on the search space (phenotype). The implications
of such an ansatz are far-reaching: An exploration density now exists
on both, the base space and the search space. In both spaces notions
as neighborhood or topology should be constituted only by the
exploration density. In this respect, the genotype-phenotype mapping
is a \emph{lift} of (topological) structure from the base space to the
search space.

To investigate the implications, we assume that the exploration
density on the base space is one of \emph{independent} random
variables. Then, for a given mapping, we investigate the exploration
density on the search space; in particular the correlations and mutual
information between phenotypic variables. This structuredness of
phenotypic exploration coherently implies notions as ``modularity''
and ``functional phenotypic complex''. Concerning the adaption of this
structure, we will argue for a self-adaptive mechanism, in place of a
statistical analysis of characters of good solutions (as with the
Bayesian ansatz). A major goal of this paper is formal and notational
clarity of such issues.

The paper is organized as follows: The next section starts by
introducing a general notation of evolutionary search. This notation
emphasizes the role of the exploration density in the search space
and, even more, the way of parameterization of this density. We call
the latter ``exploration model''. An important point of this section
is that most evolutionary algorithms differ just by this exploration
model. Since it distracts from the major line of this paper we moved
this reinvestigation of existing evolutionary algorithms to appendix
\ref{ExpMod}.

In section \ref{IndMod} we introduce and formalize the idea of
indirect modeling. Instead of parameterizing the exploration density
directly on the search space, we introduce the additional base space,
parameterize a simple density thereon, and lift this density on the
search space. We compare this to the lift of a topological or metrical
structure onto a manifold from a simple-structured base space. Notions
as pleiotropy and functional phenotypic complex are discussed as
implications of such a lift. We also relate the biological view on
indirect modeling (here, via a genotype-phenotype mapping) and
adaptive exploration to our formalism.

Section \ref{NeuBas} begins by reflecting and criticizing the usual
definition of self-adaptability. We introduce a new definition which
is based on the notion of neutrality: Different genotypes that map to
the same phenotype may represent (also topologically) different
exploration densities. Thus, such genotypes may represent very
different information and neutrality is not necessarily a form of
redundancy as is often claimed. By this definition, neutrality becomes
a central concern and we briefly review other research on this subject
in order to argue for the plausibility of our interpretation.

Finally, in section \ref{ExaSel} we exemplify all these concepts with
a running system. Simulations show that the exploration density adapts
to the problem structure by (self-adaptive) walks on neutral sets. In
particular, the pair-wise mutual information between phenotypic
variables resembles the modularity of the fitness function. We also
observe and explain a drift towards short representations. The
experiments are based on a grammar-type recursive encoding which is
thoroughly motivated by the previously developed concepts.

\section{The central role of the exploration model}\label{ForMod}

The goal of this section is to show that the central concern of
evolutionary search, esp. evolutionary algorithms, is the modeling of
exploration. We will show that the main difference between specific
evolutionary algorithms is their ansatz to model exploration.

\begin{figure}[t]\center
\input{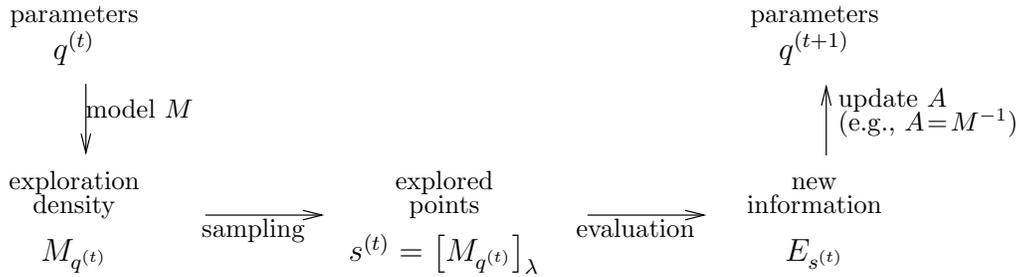}
\caption{The general scheme of evolutionary search.}
\label{evoalg}
\end{figure}

Perhaps the most general idea of stochastic search, global random
search, is described by \citeN{zhigljavsky:91}. The formal scheme of
global random search reads:
\begin{enumerate}
\item Choose a probability distribution on the search space $P$.

\item Obtain points $s_1^{(t)},\dots,s_{\lambda}^{(t)}$ by sampling
  $\lambda$ times from this distribution. Evaluate the quality of
  these points.

\item According to a fixed (algorithm dependent) rule construct a new
  probability distribution on $P$.

\item Check some appropriate stopping condition; if the algorithm is
  not terminated, then substitute $t \gets t+1$ and return to step
  (ii).
\end{enumerate}
This concept is general enough to include also evolutionary
algorithms. However, the formulation lacks to stress that the
exploration density needs to be parameterized (and instead stresses
the choice of update rule in step (iii)). We will stress the
parameterization of exploration densities and call it the
\emph{exploration model}. It is this model that we focus on. We now
formalize evolutionary search in analogy to global random search, but
with different focus:

In general we assume that the task is to find an element $p$ in a
search space $P$ which is ``superior'' to all other points in
$P$. Here, superiority is defined in terms of a quality measure for
the search problem at hand (usually a fitness function). If $P$ is too
large to evaluate the quality of all $p \in P$, the strategy is to
explore only a few points $(p_1,..,p_\l)$, evaluate their quality, and
then try to extract information on where to perform further
explorations. We capture this view on evolutionary search in an
abstract formalism that is capable to unify the different specific
approaches. Below, we exemplify each step of the scheme by embedding
the Simple Genetic Algorithm (SGA) \cite{vose:99} in the
formalism. See also figure
\ref{evoalg}.

\newcommand{\exam}[1]{{\tiny\rm }}

\begin{definition}(Evolutionary exploration)
\vspace{-\topsep}\vspace{-\parskip}
\begin{enumerate}
\item The only information maintained for evolutionary search is a
  finite set of parameters $q^{(t)} \in Q$ that uniquely define an
  \intro{exploration density} $M_{q^{(t)}}$ on $P$. Here, we call $M$
  the \intro{exploration model}, actually a map from $Q$ to the space
  $\L$ of densities over $P$. In general, the variety $M_Q=\{M_q\, |\,
  q \in Q \}$ of representable densities is limited.

\item Given some parameters $q^{(t)}$, exploration starts by choosing
  $\l$ samples $s^{(t)}_{i=1..\l}$ of the exploration density. We use
  brackets to indicate this sampling:
\begin{align}
s^{(t)}=[M_{q^{(t)}}]_\l \in P^\l \;.
\end{align}
  Here and in the following, we disregard the possibility of elitists.
  Taking them into account would require to append selected points
  $(p_1,..,p_\m)$ of $P$ to $s^{(t)}$,
\begin{align}
s^{(t)}=[M_{q^{(t)}}]_\l \oplus (p_1,..,p_\m) \in P^{\m+\l} \;.
\end{align}

\item We require the existence of an \intro{evaluation} $E:\, P^\l \to
  \L$ which maps the exploration sample $(s_1,..,s_\l)$ to a density
  over $P$ with support $\{s_1,..,s_\l\}$. This evaluation is applied
  to our exploration points:
\begin{align}
E_{s^{(t)}} = E\big( [M_{q^{(t)}}]_\l \big) \in \L \;.
\end{align}
  One should interpret $E$ as ``density of quality'' rather than a
  probability density.

\item Finally, there exists an \intro{update operator}
\begin{align}
A:\, q^{(t)} \times E_{s^{(t)}} \mapsto q^{(t+1)} \;.
\end{align}
  In general, this operator is supposed to exploit the information in
  $E_{s^{(t)}}$.
\end{enumerate}
\end{definition}

\paragraph{Example} (The Simple Genetic Algorithm)
\vspace{-\topsep}\vspace{-\parskip}
\begin{enumerate}
\item
  The SGA (without crossover) is a typical example of
  \intro{population-based modeling}: $q^{(t)} = (p_1,..,p_\m) \in
  P^\m$ is a discrete population and $\tilde M_{p_i}$ specifies the
  offspring density for each single individual. We call $\tilde
  M_{p_i}$ \intro{exploration kernels}. The total exploration density
  reads
\begin{align}
M_q=\frac{1}{\m} \sum_{i=1}^\m \tilde M_{p_i} \;.
\end{align}
  We note that the key feature of population-based modeling is its
  capacity to represent multi-modal exploration densities.

\item
  In the SGA, $s^{(t)}$ are new offsprings. The algorithms does
  not explicitly construct the complete exploration density
  $M_{q^{(t)}}$; rather, the drawing of mutations for each individual
  resembles a sampling of the exploration kernels.

\item
  For the SGA, evaluation is proportional to a given fitness
  function.

\item
  The update rule of the SGA can be written as
\begin{align}
q^{(t+1)} = \Big[ E\big( [M_{q^{(t)}}]_n \big) \Big]_n \;.
\end{align}
  In words: From the parent population $q^{(t)}$ generate $n$
  offsprings $[M_{q^{(t)}}]_n$, evaluate their fitness and select $n$
  new individuals by sampling their evaluation.

\end{enumerate}

One might assume that evolutionary algorithms mostly differ with
respect to the update rule. However, we claim that the choice of the
exploration model is crucial and that, given such a model, two generic
update operators are canonical and widely in use:

\begin{definition}(Adopting and approaching updates)
\begin{block}
  It is the \intro{adopting update} to choose the update operator such
  that $M_{q^{(t+1)}}$ is a best possible approximation of
  $E_{s^{(t)}}$ within the model class $M_Q$ (with respect to some
  chosen metric $D$ on $\L$):
\begin{align}
q^{(t+1)} = \mathrm{arg}\underset{q}{\mathrm{min}}\, D(M_q \!:\! E_{s^{(t)}}) \;.
\end{align}
  We will abbreviate this formula by using the simplified notation
  $A=M^{-1}$:
\begin{align}
q^{(t+1)} = M^{-1}(E_{s^{(t)}}) \;.
\end{align}

  Second, many algorithms realize not an adopting but rather an
  \intro{approaching update} by slowly adapting $q^{(t)}$. Here, the
  parameters must be continuous. The generic update rule reads
\begin{align}
q^{(t+1)} = (1-\a)\; q^{(t)} + \a\; M^{-1}(E_{s^{(t)}}) \;,
\end{align}
  for some constant $\a \in [0,1]$.
\end{block}
\end{definition}

\paragraph{Example} (Update operator of the SGA)
\begin{block}
  The update operator of the SGA is strongly related to the adopting
  update: The sampling $[E_{s^{(t)}}]_n$ of the evaluation density can
  be interpreted as ``finding new parameters $q^{(t+1)}$ that
  approximate $E_{s^{(t)}}$ in the population-based model''. The
  quality of this approximation is reflected by the sampling error.
\end{block}

\bigskip

Both of these canonical update operators are derived from
$M^{-1}$. Thus, when we show that most existing evolutionary
algorithms realize these operators, then we stress the importance of
the choice of exploration model. Note that any algorithm, when
embedded in the upper formalism, is uniquely characterized by the
choice of model $M$, the update operator $A$ (eventually derived form
$M$), the evaluation $E$ (given at hand) and the sampling size $\l$.

It is, of course, possible to think of exceptions that cannot be
embedded in this formalism. However, in appendix \ref{ExpMod} we show
how the formalism allows an embedding of -- and a unified view on --
very different state-of-the-art evolutionary algorithms. Indeed, those
evolutionary algorithms mainly differ with respect to their
exploration model.

\section{An indirect model of exploration}\label{IndMod}

After we stressed the importance of exploration modeling we
concentrate on the specific case of modeling defined as follows:

\begin{definition}(Indirect exploration modeling)
\begin{block}{}
  To model an exploration density over $P$, introduce a \intro{base
    space} $G=X^n$ and a \intro{base density} $M^G_q$ over $G$ such
  that the variables $x \in X$ are independent with respect to
  $M^G_q$. Then, introduce a \intro{GP-map} $h:\, G \to P$ that
  induces the exploration density $M_q=M^G_q \circ h^{-1}$ over $P$.
  Here, $h^{-1}(p) \subset G$ is a subspace of $G$ called
  \intro{neutral space} of $p \in P$; and $M^G_q\big[h^{-1}(p)\big]$
  is evaluated via integration. The class of allowed GP-maps and base
  densities limits this model $M$. The triplet $(G,h,M^G)$ is also
  referred to as \intro{coding}.
\end{block}
\end{definition}

In the following, in order to refer to their biological
interpretations, we will also use the names \intro{phenotype space}
for the search space $P$, \intro{genotype space} for the base space
$G$, and \intro{phenotype-genotype mapping} for $h$.

Also, we call the independent variables $x \in X$ \intro{genes} and
say \intro{we introduce genes on $P$} when introducing such a GP-map
and stressing the introducing of a representation via
\emph{independent} variables. This can be seen in analogy to the
introduction of local coordinates on a manifold by a local map from a
base space of (Cartesian) variables. There is, however, a crucial
difference: The map $h$ does not need to be one-to-one. If $h$ is
non-injective, there exist different genotypes $g_i$ that map to the
same phenotype. Then there exist different neighborhoods $U_{g_i}$
that map to eventually \emph{different} neighborhoods of the
\emph{same} phenotype. This change of neighborhood is of major
interest. It allows a variability of exploration. The next section
will address this important issue in detail.

As an example for indirect modeling, note that the CMA (see appendix
\ref{ExpMod}) may be interpreted as indirect modeling: it restricts
the class of GP-maps to affine transformations; the translational part
is encoded in the population's center of mass and the linear part is
encoded in the covariance matrix; the base space is $G=\RRR^n$ with
normal density $\NN(0,1)$.

\subsection{Characters of indirect exploration: Pleiotropy, 
mutual information, lift of topology, neutrality}\label{ChaExp}

The introduction of a GP-map leads to some straightforward definitions
and notions. We use this section to briefly introduce some.

\paragraph{Pleiotropy.}
In a biological context one may define pleiotropy as ``the phenomenon
of one gene being responsible for or affecting more than one
phenotypic characteristic''. Our previous definitions allow to
translate this notion into our formalism: Genes are independent (with
respect to the base density) variables of $G$. One gene affecting more
than one variable of $P$ means that the change of one variable in $G$
leads to the change of many variables in $P$. Thus pleiotropy means
that the base density of independent variables is mapped on an
exploration density of non-independent variables; pleiotropy may be
measured by the correlatedness of variables of $P$ with respect to the
exploration density. We refer to this also as structure of the
exploration density. In particular, we will measure pleiotropy as the
mutual information contained in the exploration density.

\paragraph{Population-based indirect modeling.}
Population-based modeling was defined in section \ref{ForMod}. We
briefly clarify notations in the indirect modeling case: The
parameters $q \in Q$ are a population $(g_1,..,g_\m) \in G^\m$ on the
base space and the exploration kernels $\tilde M^G_{g_i}$ are such
that the total exploration density reads:
\begin{align}
M_q&=M^G_q \circ h^{-1}
= \Big[ \frac{1}{\m} \sum_{i=1}^\m \tilde M^G_{g_i} \Big] \circ h^{-1}
= \frac{1}{\m} \sum_{i=1}^\m \Big[ \tilde M^G_{g_i} \circ h^{-1} \Big]
=: \frac{1}{\m} \sum_{i=1}^\m \tilde M_{g_i} \;.
\end{align}

\paragraph{Lift of topology.}
For population-based modeling, the exploration kernels associate a
density of offsprings to each individual. Form a topological point of
view, this defines a neighborhood (of most probable offsprings) for
each individual, referred to as variational topology.

In the case of indirect modeling, the kernels $\tilde M^G_{g}$ on the
base space are lifted to kernels $\tilde M_{g} = \tilde M^G_{g} \circ
h^{-1}$ on the search space. This means a lift of topology.

\paragraph{Neutrality.}
The possibility of a non-injective GP-map $h$ automatically leads to
the definition of neutrality.\footnote{More precisely, if also
  considering a fitness function $f:\, P \to \RRR$, we denote
  non-injectiveness of $h$ by \intro{phenotypic neutrality} and
  non-injectiveness of $f$ with \intro{fitness neutrality}. In this
  paper, only phenotypic neutrality will be addressed to.} In
particular we define $h^{-1}(p)$ as the neutral set of $p \in P$.
Further, the neutral degree of $g \in G$ is defined as the probability
\begin{align}
\tilde M_g [h(g)] = \tilde M^G_g [h^{-1}\!\circ\!h(g)] \;.
\end{align}
This reads: Take some individual $g \in G$ and let
$N=h^{-1}\!\circ\!h(g)$ be the neutral space ``around'' $g$. Now
measure the probability $\tilde M^G_g[N]$ for landing in this neutral
set when exploring from $g$.

Such measures are thoroughly discussed by \citeN{schuster:96} and
\citeN{fontana:98} (see also section \ref{IntNeu}). However, in these
publications, the variational topology rather than the probability is
emphasized. For completeness we append: Let neighborhoods be defined
in $G$ and let $B_r(g)$ be the $r$-ball around $g$ in $G$ (those
points linked to $G$ by at least one chain of no more than $r$
neighbors). We call the maximal connected component $N_g \subset
h^{-1}\!\circ\!h(g)$ with $g \in N_g$ neutral network of $g \in G$ and
define:
\begin{align}
& |h^{-1}(h(g)) \cap B_1(g)|
&& \text{neutral degree}\text{ of $g \in G$}
\end{align}

\subsection{Indirect exploration modeling in biology}\label{WagAlt}

One may argue that algorithms as discussed in appendix \ref{ExpMod}
are hardly plausible in nature and thus without relevance for
biology. What mechanisms should keep track of dependencies in nature,
model distributions by storing a Bayesian network or a covariance
matrix, and how should such knowledge be taken into account when
creating new offsprings?

Nevertheless, a biologist may in principle ask the same questions; we
refer to \citeN{wagner:96}: How comes that some phenotypic characters
are obviously correlated and others are not? How comes that a single
gene in Drosophila can trigger the expression of many others and
thereby the growth of a whole eye at different places on the body?
The existence of pleiotropy is obvious; are its specific mechanisms an
accident, an unavoidability, or the result of evolutionary
optimization? What is optimized when adapting pleiotropy?

The idea of Wagner and Altenberg is that in nature the
genotype-phenotype mapping is adaptable and does adapt in such a way
that pleiotropy between independent phenotypic characters is decreased
(in order to allow for an unbiased, parallel search) while pleiotropy
between correlated phenotypic characters may increase (in order to
stabilize the optimal \emph{relative} value of these characters). For
example, pleiotropy between the existence of the eye's cornea and its
photoreceptors is high because one alone won't contribute to selection
probability without the other. In contrast, pleiotropy between
characters of the immune system is low in order to allow a fast,
parallel optimization of different protection mechanisms which each
separately contribute to selection probability. We mimic a discussion
by picking some quotations of \citeN{wagner:96} and adding a comment:

\paragraph{Concerning evolvability}
\begin{block}
  ``Evolvability is the genome's ability to produce adaptive variants
  when acted upon by the genetic system.'' \refer{sec 5, par 2}
\end{block}
In our words: Evolvability denotes the capability of a system to model
a desired exploration distribution.

\begin{block}
  ``The thesis of this essay is, that the genotype-phenotype map is
  under genetic control and therefore evolvable.'' \refer{sec 2, par
    9}
\end{block}
In the case of indirect modeling, the GP-map induces the exploration
density on $P$. Concluding, though, that evolvability requires a
GP-map being ``under genetic control'' is questionable from our point
of view. We reflect this circumstance in detail in the section
\ref{NeuBas}.

\paragraph{Concerning modularity}
\begin{block}
  ``Modularity is one example of variational property.'' \refer{sec 1,
    par 3}
\end{block}
Modularity is a property of the exploration density. It denotes
correlations, i.e.\ mutual information, between variables of $P$. We
discussed such correlations in section \ref{ChaExp} in the context of
pleiotropy and structure of exploration.

\paragraph{Concerning functional phenotypic complexes}
\begin{block}
  ``The key feature is that, on average, further improvements in one
  part of the system must not compromise past achievements.''
  \refer{sec 5, par 10}

  ``By modularity we mean a genotype-phenotype map in which there are
  few pleiotropic effects among characters serving different
  functions, with pleiotropic effects falling mainly among characters
  that are part of a single functional complex.'' \refer{abstract}

  ``Independent genetic representation of functionally distinct
  character complexes can be described as modularity of the
  genotype-phenotype map.'' \refer{sec 6, par 1}

  ``Evolution of complex adaptation requires a match between the
  functional relationships of the phenotypic characters and their
  genetic representation.'' \refer{sec 6, par 6}
\end{block}
In essence, the exploration density should have the character that
some variables in $P$ are mutually independent while others are
dependent. Reflecting that adaptation can only occur by extracting
information from the evaluation density $E_s$ we claim that the notion
of a ``functional complex'' or a ``functionally distinct [phenotypic]
character complex'' may \emph{only} be constituted via this evaluation
density $E_s$. More precisely, we define a \intro{functional
phenotypic complex} as a set of variables of $P$ that are highly
dependent on each other (with high mutual information) but only weakly
dependent on other phenotypic characters --- all with respect to the
evaluation density $E_s$. The ``required match'' between these
properties of the exploration distribution and the evaluation
distribution motivates the adopting or approaching update as
introduced above.

\section{Neutrality as basis of self-adaptability of
exploration}\label{NeuBas}

So far, we stressed the importance of exploration modeling and focused
on the special case of indirect modeling. We did not yet address the
problem of how the exploration density can be adapted in the indirect
modeling case. This section gives an answer by providing a strict
definition of self-adaptability, which considers neutrality as a key
feature. We will also review other interpretations of neutrality and
argue in favor of our interpretation.

Obviously, if exploration is described by means of fixed kernels
around the positions of individuals, the exploration density varies
when individuals move on. But this does not quite capture what we
actually meant by requiring variable exploration. Rather it is
intuitive to call for ``adaptive codings''. The review \cite{eiben:99}
(and also \cite{smith:97}) summarizes and classifies such
approaches. Their discussion is based on the assumption that the
coding $(G,h,M^G)$ depends on some parameters $x \in X$ called
\intro{strategy parameters}; we write $(G_x,h_x,M^G_x)$. They classify
different approaches by distinguishing between different choices of
$X$:
\begin{enumerate}
\item $X$ are parameters altered by some deterministic rule (e.g.,
  function in time) independent of any feedback from the evolutionary
  process. (\intro{deterministic})

\item $X$ are parameters depending on feedback from the evolutionary
  process. (\intro{adaptive})

\item $X$ is part of the genotype. (\intro{self-adaptive})
\end{enumerate}
Option (i) is of no interest here. It is very important to distinguish
between (ii) and (iii). Option (ii) means to analyze the evolutionary
process, namely the evaluation density and the exploration density
itself, and deterministically deduce an adaptation. Good examples are
the algorithms presented in appendix \ref{ExpMod}. Option (iii) means
that adaptation becomes a stochastic search itself --- the search for
a good exploration density is itself determined by just this
exploration.

However, as formulated above, following option (iii) is quite
irritating since, after adding some strategy parameters $X$ to $G$,
the GP-map $h$ still maps $G \to P$ and it is formally incorrect to
think of $h$ as being parameterized by variables of $G$. One might
want to escape this circle by splitting $G$ into two parts, the
strategy part $X$ and the objective part $\tilde G$, $G=\tilde G
\times X$. Then, for some strategy parameters $x \in X$, one may
define $h:\, \tilde G
\times X \to P$, $(g,x) \mapsto h_x(g)$ and call $h_x$ an adaptive
GP-map. However, in general it is unclear which part of $G$ is to be
considered as strategy part and which as objective. Only in some
cases, e.g.\ if simply adding control parameters that have no direct
effect on the phenotype (neutral parameters!), this splitting seems to
be straightforward. Also, one could argue that the mutation rate of
the strategy part is kept very low. Formally and conceptually, though,
these arguments are unsatisfactory and thus we reject the definition
of self-adaptability as given by option (iii). Instead, we circumvent
such problems by defining:

\begin{definition}(Self-adaptive exploration)
\begin{block}{}
  Given an indirect, population-based model $M$ with GP-map $h$,
  exploration at $x \in P$ is defined \intro{self-adaptable} if the
  exploration kernel $\tilde M_g=\tilde M^G_g \circ h^{-1}$ varies for
  different $g \in h^{-1}(x)$ in the neutral set of $x$. The variety
  $\{\tilde M_g \,|\, g \in h^{-1}(x) \}$ of different exploration
  kernels represents the scope of self-adaptability.
\end{block}
\end{definition}

\begin{figure}[t]\hspace{-10mm}
\input{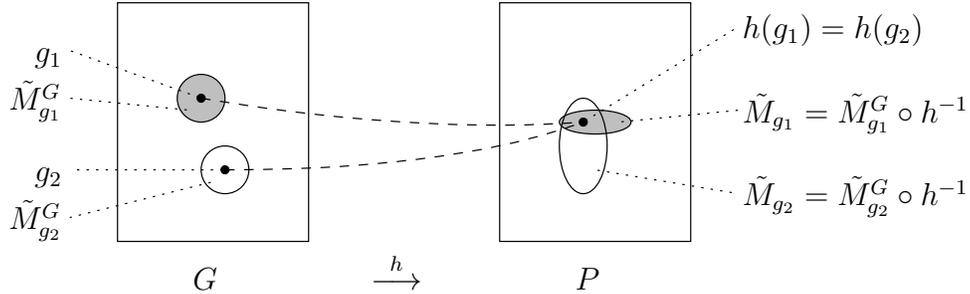}
\caption{
  Two different points $g_1,g_2$ in $G$ are mapped onto the same point
  in $P$. The elliptic ranges around the points illustrate the
  exploration kernels by suggesting the range of probable mutants.
  Thus, the two points $g_1,g_2$ belong to one neutral set but
  represent two different exploration strategies.}
\label{strategy}
\end{figure}

What does this definition mean? Assume that one individual $g \in G$
is drifting in a neutral set $h^{-1}(x)$. Meanwhile, although its
image $h(g)$ is not changing at all, the probability distribution of
\emph{offsprings} in $P$ (i.e.\ the exploration kernel $\tilde M_g$
associated to it) may change very well. This is how the definition
captures the ability of exploration to adapt. See figure
\ref{strategy} for an illustration.

As a simple example we note that adding (neutral) mutation rate
parameters aligns with this definition: Changing such strategy
parameters actually is a neutral walk but varies the exploration
kernels (e.g.\ by resizing them). Such and similar methods, may be
understood as ``local rescalings of neighborhood in $P$''; distances
(probabilities to reach neighbors within one generation) are
rescaled. However, such methods do not aim at varying the variational
topology within $P$: the probabilities for mutations into the
neighborhood change, the neighborhood itself though is not varied.
The generality of our definition also captures the latter kind of
variability and it will be a major goal of this paper to exemplify it
by introducing neutral variations that do vary the variational
topology on $P$.

In the following we will exclusively focus on self-adaptability of
exploration as defined above.

\paragraph{Note:}
Focusing only on self-adaptability (neglecting option (ii)), we want
to emphasize that we always consider the GP-map $h$ to be fix, i.e.\ 
non-varying during evolution --- and that this is \emph{not} a
restriction, \emph{not} a loss of generality. If one would protest and
claim that $h$ should be variable by depending on genes in $G$, we
veto by stating that the formalism requires to collect \emph{all}
genetic parameters in the space $G$, that by definition the GP-map $h$
is the map which maps \emph{all} $G$ on $P$, and thus it is formally
incorrect to speak of $h$ as depending on genes in $G$.

Of course, others may have another point of view and this does not
diminish the profound meaning of, e.g., Wagner and Altenberg's
statement that ``the genotype-phenotype map is under genetic control
and therefore evolvable.'' \refer{sec 2, par 9} --- though from our
point of view a questionable formulation.

\subsection{Interpretations of neutrality}\label{IntNeu}

It is intuitive to believe that every little detail in nature fulfills
``some purpose''; evolution would abandon all useless mechanisms and
redundancies. The existence of something like neutrality in nature
offends this intuition: A typical example is the fact that different
codons are transcribed into the same amino acid, suggesting that
certain nucleotide substitutions have no effect whatever on the
phenotype or its fitness --- they are neutral. Such issues initiated
many investigations, pioneered by Motoo Kimura's Neutral Theory
\cite{kimura:83}. In a later paper \cite{kimura:86}, he defends
his theory against the selectionists' criticism, who argued that
neutral genes would be functionless, mere noise, and thus biologically
implausible:
\begin{block}
  ``Sometimes, it is remarked that neutral alleles are by definition
  not relevant to adaptation, and therefore not biologically very
  important. I think that this is too short-sighted a view. Even if
  the so-called neutral alleles are selectively equivalent under a
  prevailing set of environmental conditions of a species, it is
  possible that some of them, when a new environmental condition is
  imposed, will become selected. Experiments suggesting this
  possibility have been reported by Dykhuizen \& Hartl (1980) who
  called attention to the possibility that neutral alleles have a
  `latent potential for selection'. I concur with them and believe
  that `neutral mutations' can be the raw material for adaptive
  evolution.''  \refer{\citeN{kimura:86}, page 345}
\end{block}
The last section gave a clear statement of how neutrality can be
understood as ``raw material for adaptive evolution''.

The interplay between neutrality and evolvability is a central topic
also in other works. \citeN{fontana:98}, when investigating neutrality
inherent in protein folding, claim that neutrality enables
discontinuous transitions in the protein's shape space (the space
$P$): ``[Transitions] can be triggered by a single point mutation only
if the rest of the sequence [point in $G$] provides the appropriate
context [neighborhood in $G$]; they are preceded by extended periods
of neutral drift.'' \refer{last but one paragraph} Their arguments
focus on the connectivity of neutral sets which can be analyzed
theoretically by percolation theory. We agree on these generic
ideas. A precondition is however that neutral sets exist and, most
important, that exploration varies along these neutral sets --- as we
captured in the above definition.

A very intriguing study of such phenomena in nature is the one by
\citeN{stephens:99}. They empirically analyze the codon bias and its
effect in HIV sequences. Codon bias means that, although there exist
several codons that code for the same amino acid (which form a neutral
set), HIV sequences exhibit a preference of which codon is used to
code for a specific amino acid. More precisely, at some places of the
sequence codons are preferred that are ``in the center of this neutral
set'' (with high neutral degree) and at other places codons are biased
to be ``on the edge of this neutral set'' (with low neutral
degree). It is clear that these two cases induce different exploration
densities; the prior case means low mutability whereas the latter
means high mutability. They go even further by giving an explanation
for these two (marginal) exploration strategies: Loci with low
mutability (trivially) cause ``more resistance to the potentially
destructive effect of mutation'', whereas loci with high mutability
might induce a ``change in a neutralization epitope which has come to
be recognized by the immune system.''  \refer{introduction, par 4}

Finally, several models of landscapes with tunable neutrality have
been proposed to theoretically investigate possible \emph{purposes} of
neutrality \cite{barnett:98,newman:98,reidys:01}.

In this paper we present a simple setup to demonstrate the dynamics in
neutral networks in appendix \ref{IllNeu}. Using Eigen's model we show
a drift towards high neutral degree, i.e.\ towards representations of
low mutability. This effect is important to understand the experiment
we present in section \ref{TwoExp}.

\section{Paradigms of self-adaptive exploration}\label{ExaSel}

The goal of this section is to exemplify the principles discussed
above by simple and transparent (artificial) systems. In order to
setup a running system we need to make some further decisions on
\begin{block}
(i) the problem (the space $P$),\\
(ii) the GP-map (including the choice of $G$),\\
(iii) the base density (population size, mutation rates on $G$, etc.),\\
(iv) the evaluation (implementation of $E$),\\
(v) the update rule $A$.
\end{block}
In the following $P$ will simply be strings over some alphabet $\AA$;
the problem is to minimize the (Hamming) distance to a given target
string. Concerning point (iv) and (v), we will use rank-based
selection, i.e.\ we evaluate proportionally to the rank of each
individual and update the population by sampling this evaluation
density. Point (ii) and (iii) need more thorough considerations:

\paragraph{A recursive, grammar-type GP-map.}
We decide to implement the GP-map as a recursive mapping. More
precisely, $h$ is representable as a composition of a single
\intro{GP-generator} $\hat h:\, G \to G$,
\begin{align}
h = \underbrace{\hat h \circ .. \circ \hat h}_{m(\cdot) \text{ times}}
~:~ G \to P \subset G ~:~ g \mapsto 
  \underbrace{\hat h \circ .. \circ \hat h}_{m(g) \text{ times}}(g) \;.
\end{align}
This inevitably requires a choice of $G$ such that $P \subset G$. The
recursion depth $m$ may depend on the point $g \in G$. Generically, we
require that each GP-generator affects (or entangles) only a few
variables within $G$. The motivation is as follows: Structuredness of
exploration, as discussed in section \ref{ChaExp} and \ref{WagAlt},
means mutual information between variables that belong to the same
phenotypic character and less mutual information else. We want the
generator to represent elementary correlating effects (e.g.\ of
interaction), i.e.\ to constitute elementary modules. For example, an
elementary correlating effect is that one character depends also on
another and a respective generator would introduce such mutual
information by mapping one independent variable onto one which depends
on other variables. An $NK$-reaction network is a basic example: the
generator (the time step transformation) entangles $K$ variables to a
new one.

Our examples will use a grammar-type recursive mapping. The space $G$
is organized as
\begin{align}
G = P \times \big[ \AA \times P \big]^r \;,
\end{align}
which means that $g$ encompasses one structure $g_0 \in P$ (called
axiom) and $r$ tuples $g_i \in \AA \times P$ (called rules). The
GP-map $h$ applies to $g \in G$ by applying all rules to $g_0$; the
symbols $l \in \AA$ in each rule (actually the lhs label of a grammar
rule) specify how to apply the rule. (The GP-generator is the single
application of one rule to the axiom.) In our examples, the recursion
depth $m$ is always fixed (so we need no terminal symbols or other
complicated mechanisms.)

Such grammar-type encodings have been investigated in many other
respects, e.g.\ by \citeN{prusinkiewicz:89} and
\citeN{prusinkiewicz:90} discussing L-systems as natural
representation of highly regular, plant-like structures; by
\citeN{kitano:90}, \citeN{gruau:95}, \citeN{lucas:95}, and
\citeN{sendhoff:98} using grammar-encodings as representation of
neural networks. However, these approaches are not based and motivated
on a discussion of self-adaptive exploration. Thus, although in most
cases the existence of neutral sets (equivalent representations) in
grammar encodings is obvious, the importance to introduce (neutral)
variations that explore these existing neutral sets and thereby
explore different explorations strategies was not recognized and
stressed. The next paragraph concerns the introduction of such
variations.

\paragraph{Neutral variations in grammar-type encodings.}
We turn to the choice of base density, i.e.\ variability on $G$. We
assume that there exist canonical mutations on $P$, namely flip (with
probability $\a$ per symbol), insertion, duplication and deletion
(with probability $\c$ per string). Since $G$ is composed of
structures of $P$ these mutations induce standard mutations on
$G$.

However, to take all the considerations of section \ref{NeuBas} into
account, we additionally introduce neutral variations on $G$. These
variations are supposed to allow for self-adaptability as defined
above, i.e.\ they should allow neutral variations that vary
exploration. In our examples we realize such variations by rule
substitutions and creations. Specifically we introduce five kinds of
variations of $g \in G$, which are likely to be neutral but need not
always to be:
\begin{block}
  (i) Pick one rule and one structure $\in P$ (any rhs or the axiom)
  within $g$; then apply the rule once to the structure.\\ (ii) Pick
  one rule and one structure; check if the rhs of the rule is part of
  the structure; if so, replace this part by applying the rule
  inversely.\\ (iii) Pick a structure and create a new rule by
  extracting a part out of the structure and replacing it by a
  symbol.\\ (iv) Delete a rule if it is never applied during
  recursion.
\end{block}
All of these variations will occur with probability $\b$ per rule (per
structure in case (iii)).

\subsection{Basic paradigm}

Let $P$ be strings of the alphabet $\{0,1,x\}$. Consider the following
two points $a,b \in G$ to represent the same point $0101$ in $P$:
\begin{align*}
& a_0 = 01x \comma a_1 = (x \mapsto 01) \;,\\
& b_0 = xx \comma  b_1 = (x \mapsto 01) \;.
\end{align*}
If we assume that the rhs of $a_1$ and $b_1$ have considerable
mutability, the exploration kernels of $a$ and $b$ are quite
different: Probable (phenotypic) mutants of $a$ are $0111, 0100,
0110$, whereas $b$ is likely to produce mutants like $1111, 0000,
1010$. The difference of these two exploration densities is of
\emph{topological} nature.

In order to enable a transition between such different strategies, the
exploration of the corresponding neutral set must be possible. In the
upper example it is easy to define a neutral mutation from $a$ to $b$:
The rule itself is to apply to the axiom. The inverse mutation
requires an application of the rule from right to left, i.e., see if
the rhs fits somewhere and substitute by the lhs. Our system
incorporates these variations.

\subsection{Two experiments: Variability of exploration and neutral
  drift}\label{TwoExp}

\begin{figure}[t]\center
\includegraphics[scale=0.3,angle=0]{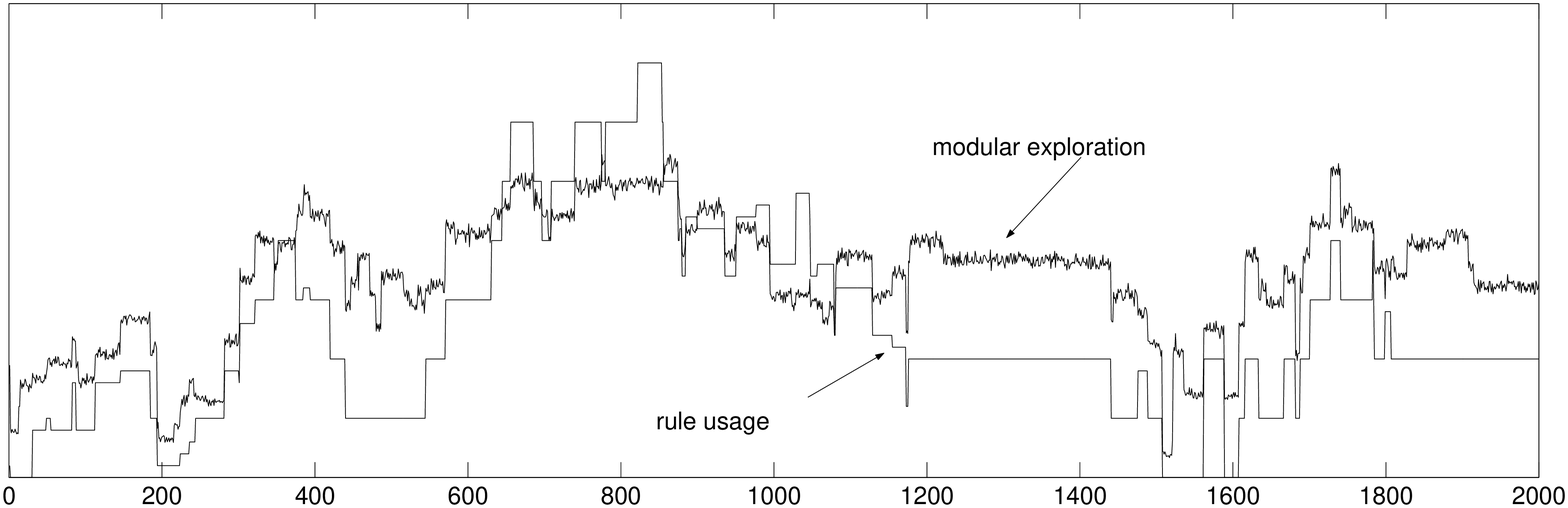}
\caption{
  A single individual is tracked when drifting on a neutral set
  spanned by neutral substitutions in its grammar-encoding. Its
  exploration density is analyzed by taking 10\,000 samples at each
  time step. 'Modular exploration' counts the probability for
  mutations that occur equally at same positions in other blocks.
  These are blocks of 5 symbols as given by the target string: 5
  $\times$ {\sc abcde}. 'Rule usage' counts how often rules are
  applied during recursion. [Population size $\m=1$; mutation
  probabilities $\a=0.001$, $\b=0.1$; recursion depth $m=10$; scaling
  of $y$-axes is only relative.] }
\label{5modules}
\end{figure}

Let $P$ be the strings over the alphabet \{{\sc a,b,c,d,e,f,g,h}\}.
The function $f$ is the Hamming distance to the fixed target string
{\sc abcdeabcdeabcdeabcdeabcde}, i.e.\ 5 times {\sc abcde}. To
demonstrate a neutral drift we consider only one individual and
initialize it with an axiom equal to the target and no rule. Selection
is (1+1), i.e.\ at each time step one offspring is produced and
selected if equally good or discarded if worse. As a result of neutral
variations, the number of rules and the probability for regular
mutations in the exploration density vary in correlation. This kind of
variability of exploration is of topological nature. The point is, we
gave an example where the topological characters of the exploration
density vary over a connected neutral set. See figure \ref{5modules}.

\begin{figure}[t]\center
\includegraphics[scale=0.3]{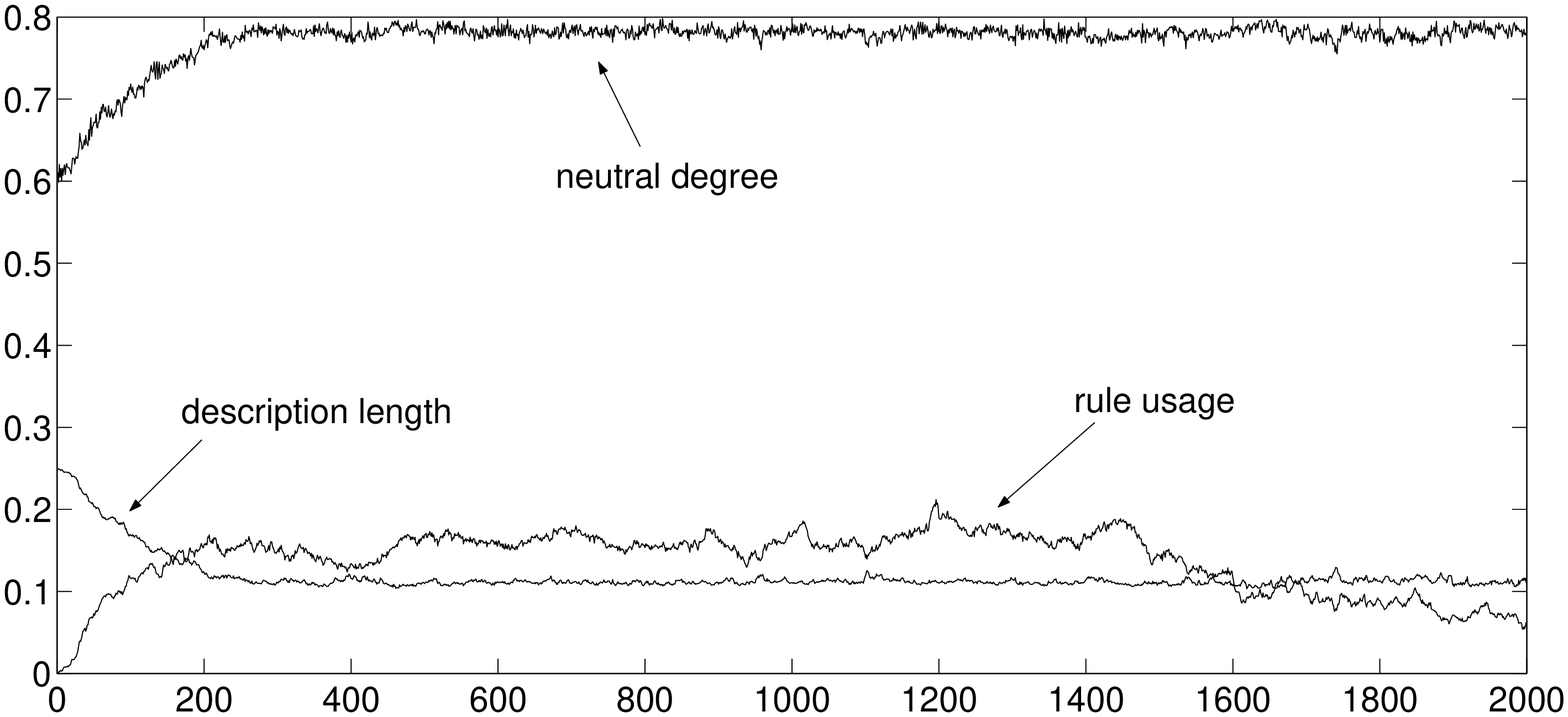}\\
\includegraphics[scale=0.3]{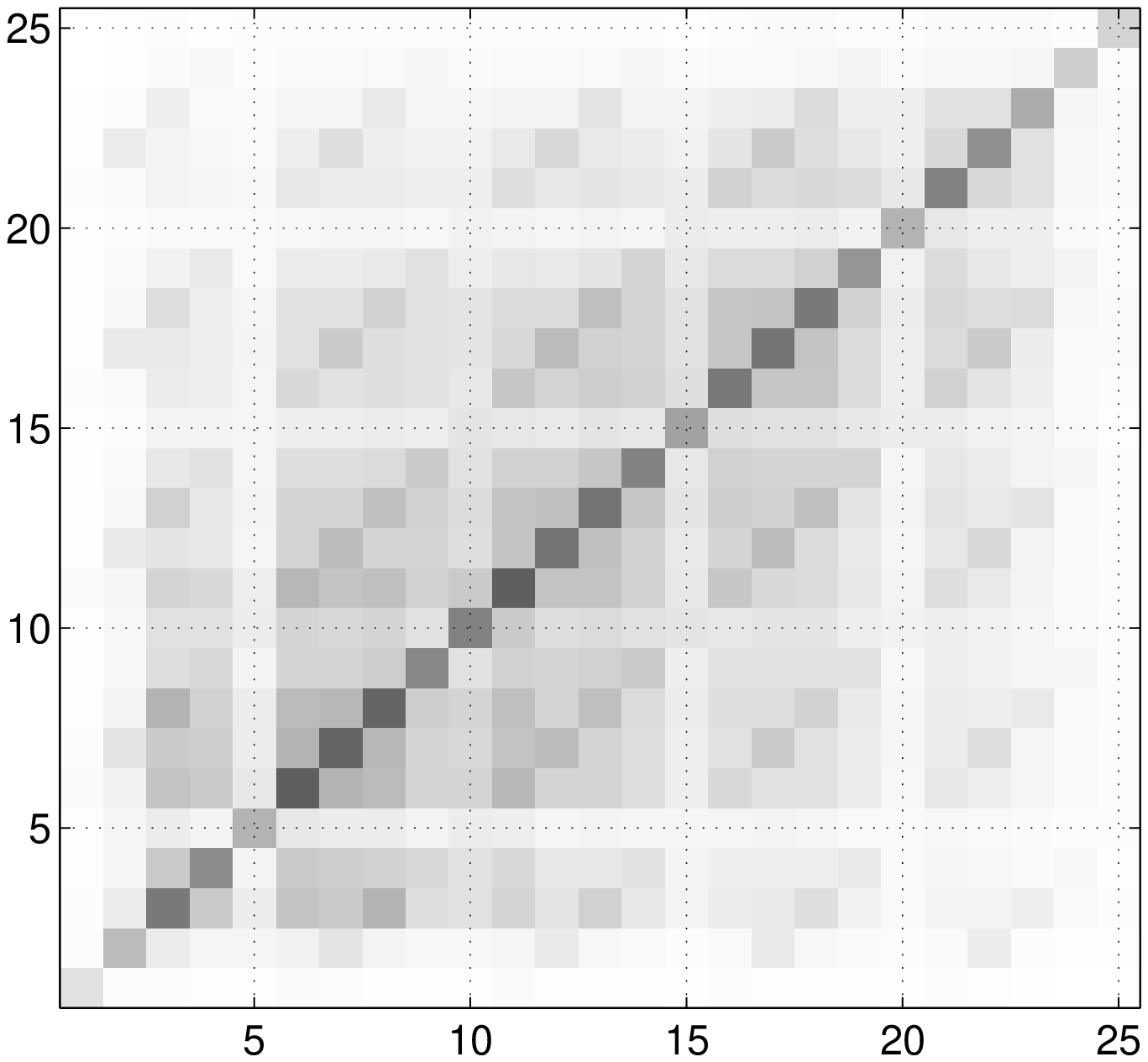}
\includegraphics[scale=0.3]{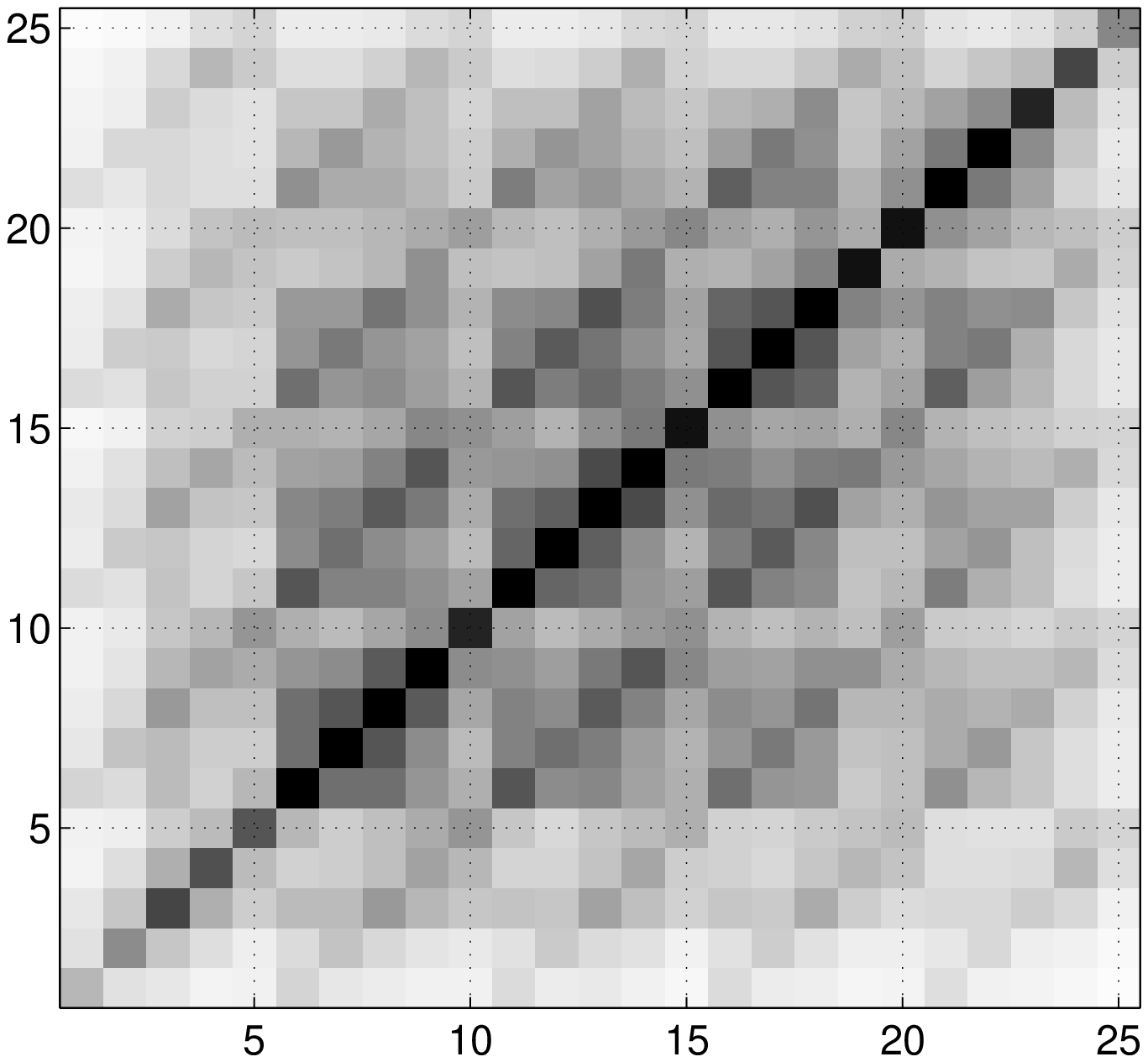}
\includegraphics[scale=0.3]{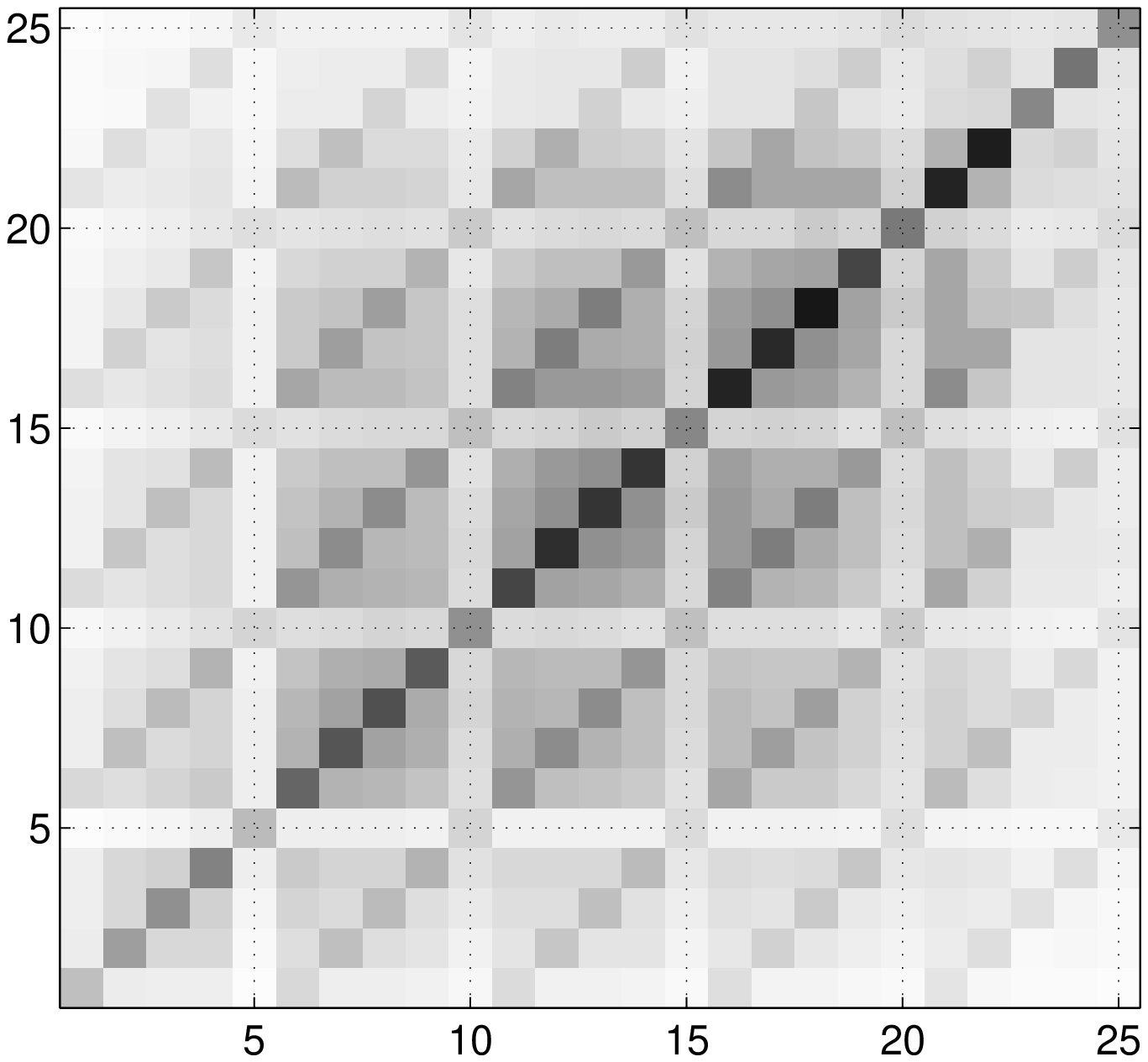}
\caption{
  Upper plot: A population is tracked when drifting on a neutral set.
  Selection is non-elitist and rank-proportional and thus pushes the
  population towards higher neutral degree. This is achieved by
  finding representations of shorter description length of the modular
  target string (5 $\times$ {\sc abcde}). (Description length equals
  the sum of the lengths of axiom and all rhs). This in turn is
  achieved by making use of rules. Lower plots: The mutual information
  between the 25 variables in the (phenotypic) exploration density is
  displayed as a matrix. The three plots correspond to times 50, 500,
  and 2000. The regular, 5-modular structure of exploration is clearly
  visible. [Population size $\m=100$; mutation probabilities
  $\a=\b=0.02$; target: 5 $\times$ {\sc abcde}; recursion depth $m=3$;
  scaling of $y$-axes is exact for neutrality, only relative for the
  rest; scaling of gray-shading is only relative.] }
\label{o4}
\end{figure}

We enhance this example by considering a population of 100 individuals
and non-elitist, rank-based selection. All individuals are initialized
as described above. The population drifts towards representations
(points in the neutral set of the target string) with high neutrality.
This effect is explained in detail in appendix \ref{IllNeu}. Here, a
high neutral degree coincides with representations of short
description length (the sum of lengths of the axiom and rhs of rules).
In order to achieve such compact representations, more rules are
extracted and included in the representation. A visualization of the
exploration density via mutual information maps exhibits its clear
structure that corresponds to the target string's structure. One may
interpret that the system has ``learned the problem's structure''. See
figure \ref{o4}.

\clearpage

\section{Conclusions}

Major parts of this paper are concerned to develop an integrated
language for evolutionary search based on the formalism of stochastic
search and emphasizing the exploration density and its
parameterization. The benefit is a unified view on different specific
approaches, their commonness and differences. For example, at first
sight it is hard to see what a CMA evolutionary strategy has in common
with the codon bias in HIV sequences. The answer is: both of them are
concerned to model the variability of future offsprings, the
exploration density; both of them by using a kind of
genotype-phenotype mapping (an affine transformation in the first
case). Also notions such as pleiotropy and functional phenotypic
complex can properly be defined on the basis of this language. This
allows to make contact between biological and computational research.
The functional meaning of a genotype-phenotype mapping is illuminated
by interpreting it as a lift of an exploration density and topology on
the search space. We showed that a non-injective genotype-phenotype
mapping can lift different exploration strategies, different
topologies to the same phenotype. This is the core of how we define
self-adaptability of exploration. The definition overcomes the formal
weakness of previous definitions and is as general as the language it
is based on. The definition opens a completely new view on the meaning
of neutrality.

In the experimental part of this paper we presented elementary
examples of these concepts. We illustrated the structure of
exploration by a gray-shade map of the mutual information within the
exploration density, a gray-shade map of pleiotropy. We exemplified
its variability during neutral drifts. And we demonstrated successful
self-adaptability of exploration where in the end the structure of
exploration perfectly matches the structure of the problem.

We will now discuss some further implications of the new view we have
developed in this paper:

\paragraph{(i) On modularity, structuredness, and evolvability.}
Given a system that functions well, how should one define what a
module or a functional complex is? One only observes that all parts
together work well as a whole. A common idea is that modules are
characterized by high interactivity within them. By high interactivity
we mean that there are high correlations between units during the time
of functioning. These are completely different kinds of correlations
than correlations between units in the evolutionary variability. It is
though possible to draw a link: Having units that are highly
interacting during functioning, the fitness might strongly depend on
their teamwork. If this is the case, also the evaluation density
should incorporate high correlations between the units (i.e.\ the
units form a functional phenotypic complex). Now, if the exploration
density should approximate the evaluation density, we also find these
correlations in the evolutionary variability.

Thus, when talking about modules, one should be aware of the
interrelations between these three levels of correlations: (1) during
functioning, (2) in the evaluation density, (3) in the exploration
density. Our definition of a functional phenotypic complex refers to
the 2nd level -- the evaluation density. Our hypothesis is that the
advantage of structured systems (and thus the selective pressure
towards structure) stems from the 3rd level:
\begin{block}\em
  Systems are structured, not because this is the only possible way of
  functioning, but because it is advantageous for variability.  The
  advantage of structured variability is its capability to explore by
  approximating the ``problem's structure'', the structure of the
  evaluation density.
\end{block}
This capability should be called \emph{evolvability}.

For example, parts of a system that contribute separately to fitness
should be varied and optimized in parallel without potentially
disturbing correlations; whereas parts of a system that only
contribute to fitness when they are tuned on each other should be
varied in correlation in order to preserve this tuning.

\paragraph{(ii) On redundancy and neutrality}
Neutrality is often thought of as redundancy. From our point of view,
this is very misleading.  As we pointed out in the context of
self-adaptability, although all the genotypes in a neutral set encode
the same phenotype, they may have very different exploration kernels.
Thus, such genotypes may carry different information. One cannot speak
of redundancy if different and relevant information is encoded. If,
however, genotypes in a neutral set have identical exploration kernels
(in the genotype space), then they are indeed redundant. Redundancy is
necessarily neutral, but neutrality is not necessarily redundant.

\paragraph{(iii) On compact representations}
Assume we use a Bayesian network to model the structure of
exploration. Then we will explicitly encode the correlations between
all phenotypic variables. In contrast, our second example shows how
compact representations correspond to highly structured exploration
and can be found by using recursive codings. The idea is that each
recursion introduces correlations in the variables. The neutral drift
towards high neutral degree (see appendix \ref{IllNeu}) induces a
selective pressure towards short representations.

\paragraph{(iv) On grammar-type encodings}
In grammar-type encodings, some single genotypic variables (genes)
might effectively represent whole groups of phenotypic variables.
Thus, when we model dependencies between variables, we can also model
dependencies between whole groups of phenotypic variables and not only
between single phenotypic variables as in the direct modeling ansatz.
This allows to introduce deep hierarchical dependencies in the
exploration density.

Most existing approaches to grammar encoding are motivated by the fact
that grammars can represent regular structures with short description
length. Instead, we claim that the most interesting point about
grammars is their capability to introduce structure in the
variability, as demonstrated in our examples. In order to explore
these capabilities in a self-adaptive manner, the inclusion of neutral
variations in recursive or grammar-type encodings is of crucial
importance. This point seems neglected in the literature.

We rigorously support Kimura's ``belief that `neutral mutations' can
be the raw material for adaptive evolution'' \cite{kimura:86}.

\appendix

\section[Exploration models of specific approaches]{The exploration model
  of different state-of-the-art evolutionary algorithms}\label{ExpMod}

To stress the importance of the concept of exploration modeling we
want to show that the main difference between specific evolutionary
algorithms is their ansatz to model exploration. In order to do so, we
embed specific algorithms in our formalism. In particular we chose to
analyze the CMA algorithm and three recent approaches which belong to
the class of ``probabilistic model-building genetic algorithms''
(PMBGAs), see \cite{pelikan:99}. All of these realize adaptive (but
not self-adaptive) exploration.

\paragraph{Covariance Matrix Adaptation (CMA),}
\cite{hansen:00}. The search space is continuous, $P=\RRR^n$. The CMA
algorithm maintains as parameters $q$ only one (center of mass) point
$p \in P$, the symmetric covariance matrix $C$, and some adaption rate
parameters. The exploration density $M_q$ is given by a linear
transformation (via $C$) of a Gaussian distribution around $p$. In
practise, the algorithm generates $\l$ normally distributed mutation
vectors $z_i \in \RRR^n$, transforms all of these vectors by
multiplying the matrix $C$, and adds these vectors to the center of
mass $p$ in order to generate the new $\l$ samples. After evaluation
of the samples it is updated as follows: $p$ is moved to the center of
mass of the selected samples and $C$ is adapted as
\begin{align}
C^{(t+1)} = (1-c)\; C^{(t)} + c\; z \otimes z \;.
\end{align}
Here, $c$ is some adaption constant and $z$ is the
average\footnote{More exactly an weighted average trace over time,
see \cite{hansen:00} Eq. 14.} of the selected mutations
vectors. (\citeN{hansen:00}, Eq. 15, write $z(z)^T$ instead of $z
\otimes z$). The point is that $z \otimes z$ is the unique symmetric
matrix which maps the equally distributed vector
$y=(\frac{1}{n},..,\frac{1}{n})$ to $z$. Thus, the update rule for $C$
corresponds to our generic \emph{approaching} update whereas $p$
\emph{adopts} the new center of mass.

\paragraph{Dependency tree modeling,}
\cite{baluja:97}. Here, the search space is discrete, $P=X^n$. In
their algorithm, the parameter $q$ that describes the next exploration
density is a dependency tree. Thus, the model is restricted to encode
only pair-wise dependencies between variables. At each time step, $\l$
samples are generated from this exploration density; the samples are
evaluated and the best $\m$ of them are selected. A probability
density $A$ of previously selected points is adapted by including
those newly selected ones (generically $A \gets (1-\a)\, A + \a\,
[E_{s^{(t)}}]_\m$). Then the dependency tree is updated by minimizing
the Kullback-Leibler divergence between $A$ and $M_q$. The tree's
update is an adopting since it approximates $A$, whereas $A$ itself is
updated according to an approaching update.

\paragraph{Factorized Distribution Algorithm (FDA),}
\cite{muehlenbein:99}. Again, $P=X^n$ is discrete. The parameters $q$
describe the conditional dependencies in pairs, triples, quadruples,
etc.\ of variables. (To be exact, the algorithm comprises also some
elitists.) The model is quite general but it relies on pre-fixed
knowledge on which pairs, triples, etc.\ exactly are to be
parameterized. At each time step, the dependencies within the
distribution of evaluated and selected points are calculated and
assigned to $q$. Therefore, this is an adopting update.

\paragraph{Bayesian Optimization Algorithm (BOA),}
\cite{pelikan:00}. $P=X^n$ is discrete. Here, $q$ is a general
Bayesian dependency network that explicitly encodes the exploration
density. Thus, the model is not limited in representing arbitrary
orders of correlation and it is flexible in which variables are
dependent by inserting and deleting connections in the network. After
selection, the network is recalculated in order to minimize (e.g.\ 
with a greedy algorithm) the distance (e.g.\ with respect to the
Bayesian Dirichlet Metric) between $M_q$ and the distribution of
selected. This is, except for elitists, also an adopting update.

\section{Illustrating neutral dynamics}\label{IllNeu}

As an illustration of neutral dynamics we present a simple example. We
assume that the search space $P$ is discrete and rather small,
$|P|=\l$. $\L$ denotes the space of densities over $P$, which actually
is a simplex. Parameter $q \in Q$ is such a density, $Q=\L$, and the
exploration density $M_q$ is a mutation $\tau\, q \in \L$ of this
density. This example omits sampling and thus evaluation $E:\, \L \to
\L$ directly applies to $M_q=\tau\, q$. The update rule is the
adopting:
\begin{align}
q^{(t+1)} = E\, \tau\, q^{(t)} \comma
q^{(t+1)}_i = \sum_{j,k=1}^\l E_{ij}\, \tau_{jk}\, q^{(t)}_k \comma
\end{align}
whereby we actually formulated Eigen's model (see e.g.\
\cite{eigen:89}) in our notation. Finding the eigenvectors of $E\,
\tau$ means finding a stationary population density. Their eigenvalues
describe their growth factor and the eigenvector with highest
eigenvalue will describe the final attractor --- the quasi-species. In
the presence of a neutral set $N$ (here a set of indices) we assume
that only individuals on this neutral set are evaluated positively and
without co-evolutionary (interacting) effects, i.e., $E$ is diagonal
and
\begin{align}
&E:\, \L \to \L \comma
p_i \mapsto \sum_j E_{ij}\, p_j
= E_{ii}\, p_i = \left\{
{0 \atop e_i(p)\, p_i} \quad
{i \not\in N \atop i \in N} \right. \quad.
\end{align}

\begin{figure}[t]\center\small
\begin{minipage}{30ex}
neutral set $N \subset P$:\\
\includegraphics[angle=0,scale=0.2]{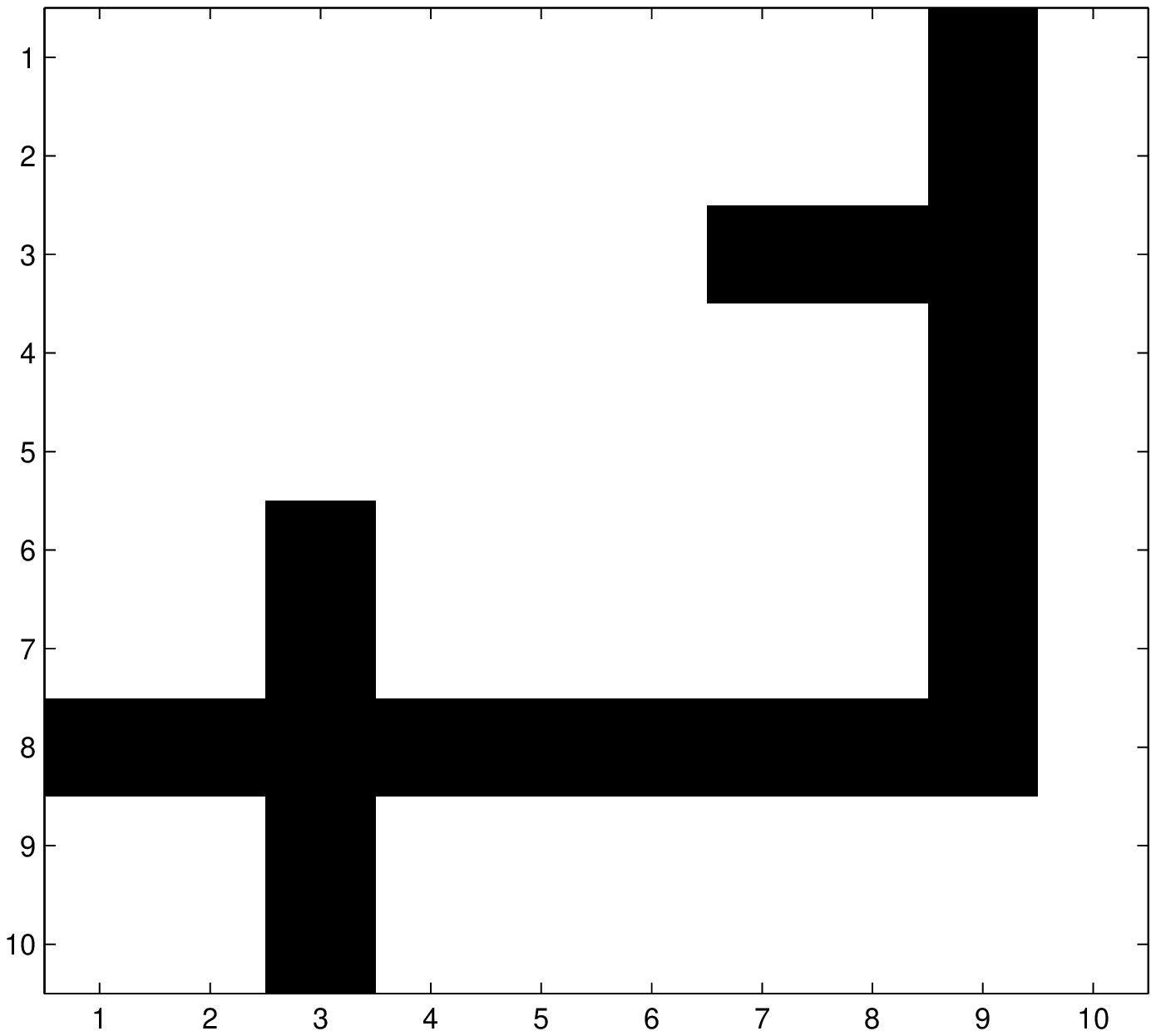}
\end{minipage}
\begin{minipage}{30ex}
First experiment\\[2ex]
density $q \in \L$:\\
\includegraphics[angle=0,scale=0.2]{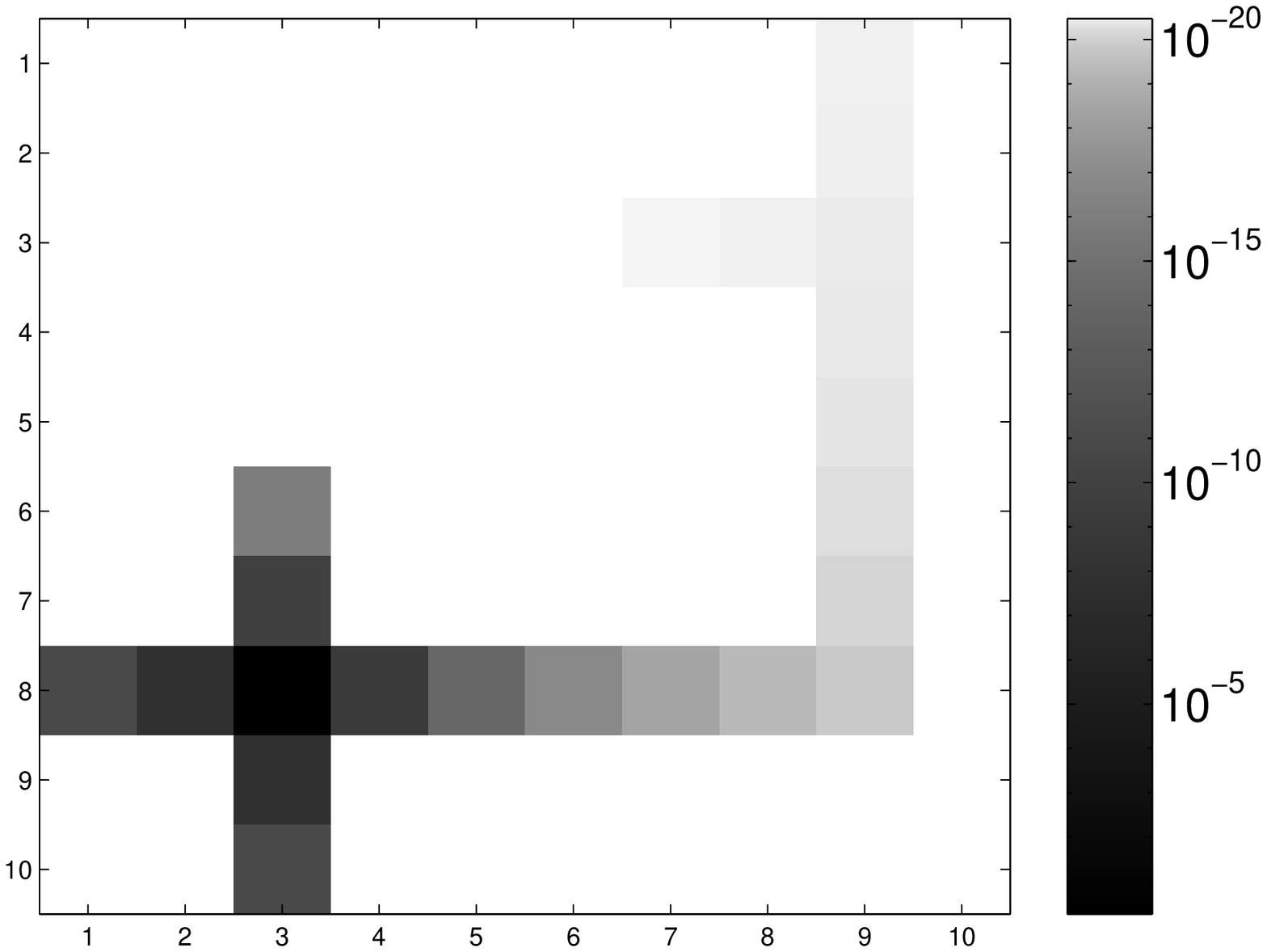}\\[2ex]
mutated density $\tau\, q \in \L$:\\
\includegraphics[angle=0,scale=0.2]{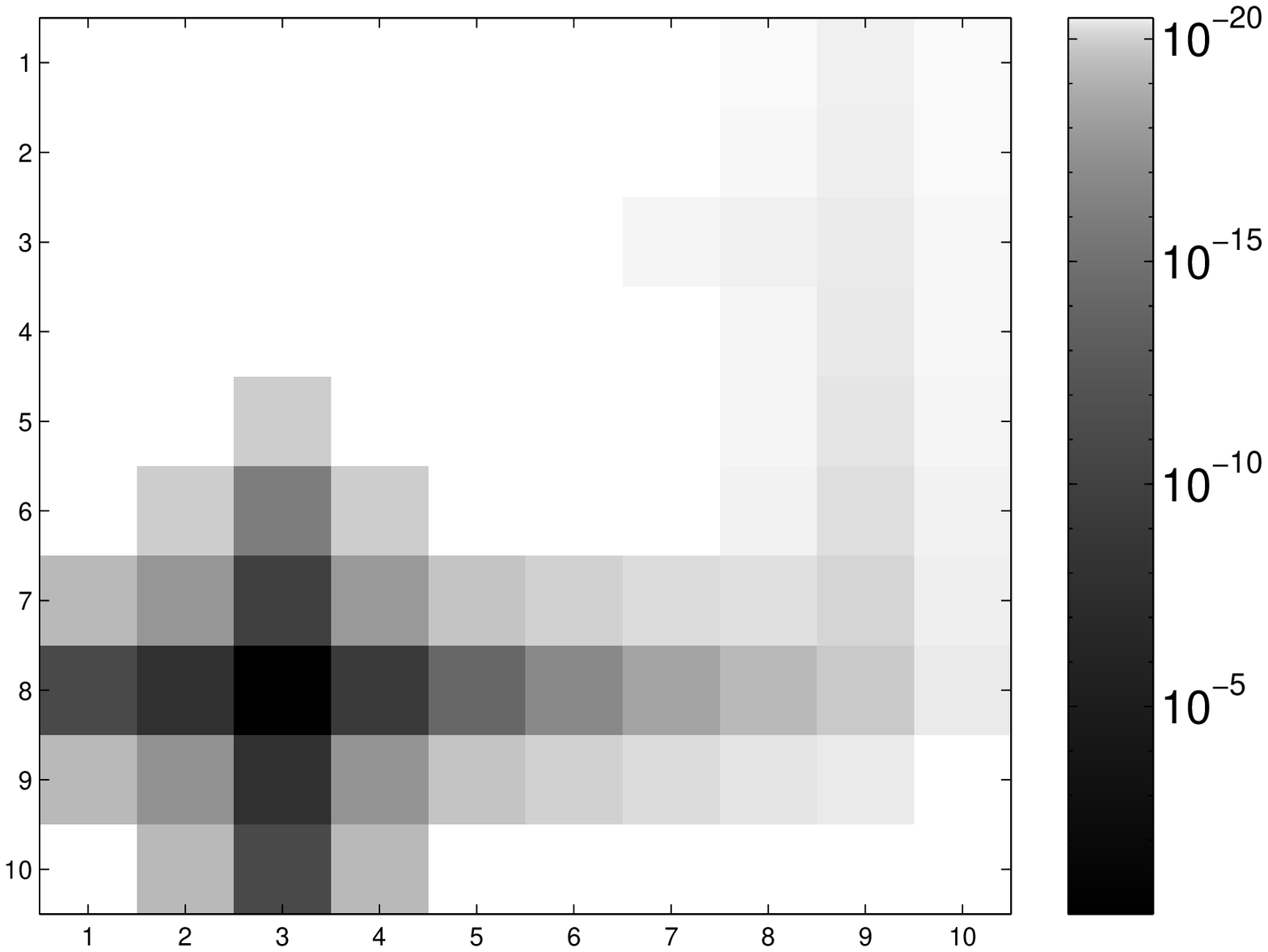}
\end{minipage}
\begin{minipage}{25ex}
Second experiment\\[2ex]
density $q \in \L$:\\
\includegraphics[angle=0,scale=0.2]{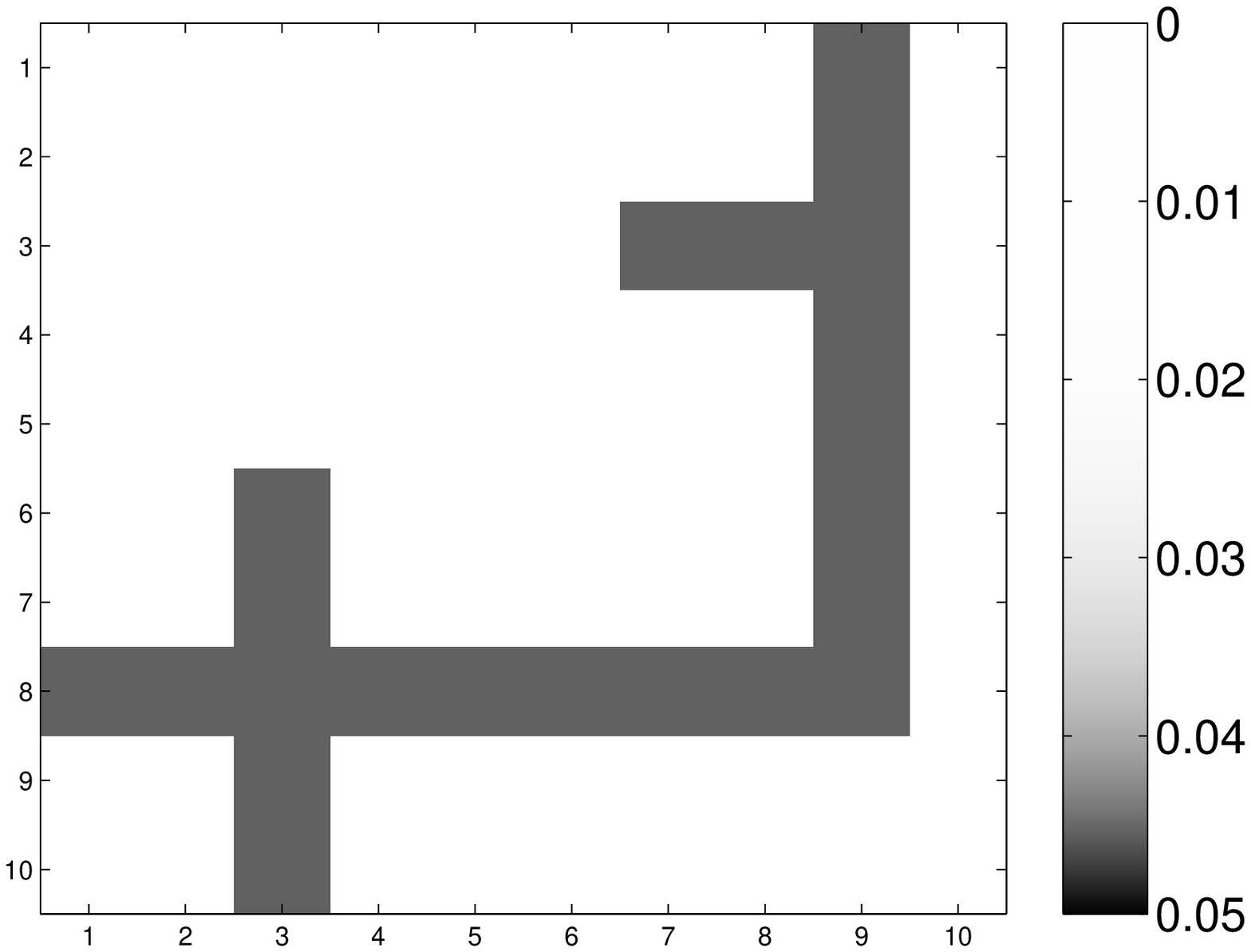}\\~\\
mutated density $\tau\, q \in \L$:\\
\includegraphics[angle=0,scale=0.2]{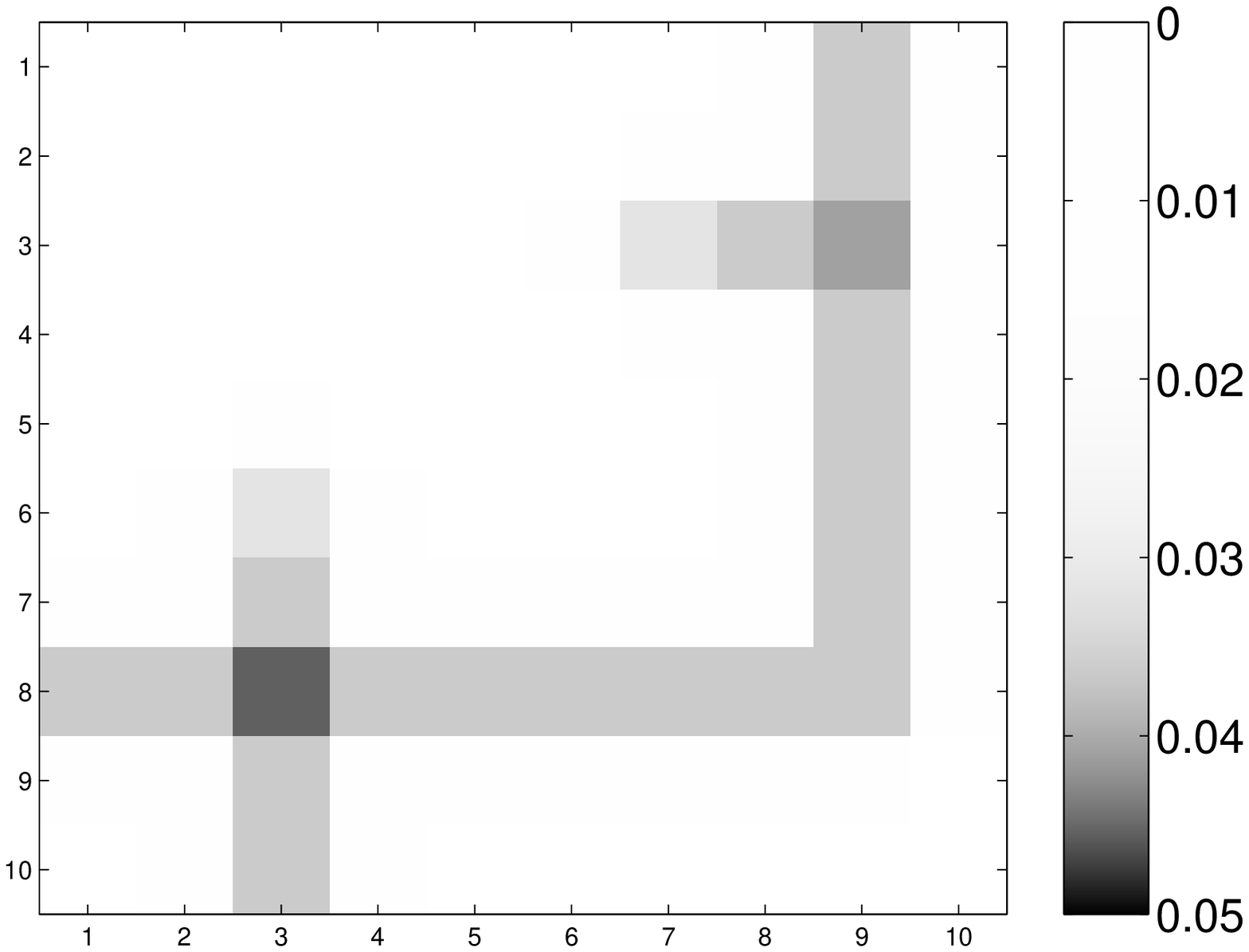}
\end{minipage}
\caption{
  The search space $P$ is represented as a 10$\times$10 board. The
  neutral set is embedded as depicted on the left. The exploration
  matrix $\tau$ corresponds to a mutation rate of $0.1$ in each of the
  four directions (up,down,right,left). In the first experiment, when
  evaluation is straightforward, e.g.\ fitness-proportional, it is
  impressive to see how strong the attraction towards the crossing
  with neutral degree $1$ (with four neutral neighbors) is. In the
  second experiment, where evaluation enforces a kind of local
  conservation of population density, the population is equally
  distributed on the neutral set, but exploration on places with high
  neutral degree is proportionally higher because they have more
  neighbors from which they ``receive'' offsprings.}
\label{neuts}
\end{figure}

We investigate two options for the evaluation factor $e_i(p)$. The
first and straightforward option is that all positions on the neutral
set are evaluated equally, then
\begin{align}
e^1_i(p)=\frac{1}{\sum_{j \in N} p_j}
\end{align}
is just the appropriate normalization factor. This option is realized
e.g.\ for fitness-pro\-por\-tion\-al evaluation (when fitness on $P
\setminus N$ vanishes) but also for fair ranking. For the second
option we enforce such positions on the neutral set with low neutral
degree --- inverse-proportionally to the neutral degree:
\begin{align}
e^2_i(p)=\frac{1/d_i}{\sum_{j \in N} (p_j/d_j)} \comma
d_i := \sum_{k \in N} \tau_{ik} \;.
\end{align}
The quantity $d_i$ is the probability for an offspring of individual
$i$ to be an element of $N$. Thus, this option increases the
evaluation of $i$ such that the probability to provide an offspring
\emph{in} $N$ becomes equal for all $i \in N$. This can be compared to
a local conservation of population density: Effectively, each parent
in $N$ will with equal probability contribute a viable offspring to
the next generation. Such a type of selection can be realized by local
selection mechanisms: From each parent produce many offsprings, let
only the best of these offsprings compete with others. As a result,
the quasi-species is simply constant on $N$ and vanishes elsewhere,
$q_{i\in N}=1/|N|$, $q_{i \not\in N}=0$:
\begin{align}
& p_i = (\tau\, q) _i = \sum_{j \in N} \tau_{ij}\, q_j = \frac{d_i}{|N|} \;, \\
& (E\, \tau\, q)_{i \in N}
= \frac{1/d_i}{\sum_{j\in N} (p_j/d_j)}\; p_i
= \frac{1}{|N|}
= q_i \;.
\end{align}
The mutated density $p_i$ is proportional to $d_i$ (which, for
individuals out of $N$, does not denote the neutral degree but rather
the probability for offsprings in $N$). Diversity is much higher than
for the first type of evaluation. See figure \ref{neuts}.

The first experiment is an explanation for the dynamics we observe in
section \ref{TwoExp}. We included the second experiment because it
realizes what one might intuitively have expected: on a neutral set
the population is distributed equally and with high diversity. We
showed what kind of evaluation one has to choose to fulfill this
expectation.

The findings are conform with Nimwegen's \citeyear{nimwegen:99} little
examples of random or selective walks on a neutral set: A blind ant
would try one (random) neighboring genotype and walk to it if it has
same fitness or stay otherwise. A myopic ant would find all neighbors
with same fitness and walk to one (random) of those. He finds that,
\emph{in temporal average}, the blind ant stays equal times at each
genotype of the neutral set whereas the myopic ant stays longer at
centers of the neutral set (i.e.\ $\propto$ the neutral degree). The
myopic ant, since it always finds a neutral neighbor, corresponds to
our second example.

\bibliographystyle{comments}
\bibliography{bibs}

\end{document}